\documentclass[10pt,conference]{IEEEtran}
\usepackage{alltt                                    
          , multirow
          , booktabs
          , listings
          , graphicx
          ,float
          ,verbatim
         ,mathtools
	,url
	,amsmath
}

\usepackage[numbers]{natbib}     
\usepackage{syntax}
\usepackage{algorithm}
\usepackage{color, colortbl}
\usepackage{enumitem}
\usepackage{threeparttable}
\usepackage{pbox}
\usepackage{balance}
\usepackage{algpseudocode}
\usepackage{hhline}
\usepackage{tikz}
\usepackage{pgf-pie}
\usepackage{framed}
\usepackage{balance}

\usetikzlibrary{arrows,shapes,backgrounds}

\usepackage{etoolbox}
\usepackage{tikz}
\usetikzlibrary{tikzmark}
\usetikzlibrary{calc}
\usepackage[all]{nowidow}

\usepackage{lipsum} 

\tikzstyle{vertex}=[ellipse,fill=black!25,minimum size=20pt, inner sep=0pt]
\tikzstyle{edge} = [draw,thin,-]
\tikzstyle{glabel} = [text width=1cm,text centered,font=\bf]
\pgfdeclarelayer{bg}    
\pgfsetlayers{bg,main}  

\usepackage{expl3}
\ExplSyntaxOn
\newcommand\latinabbrev[1]{
  \peek_meaning:NTF . {
    #1\@}%
  { \peek_catcode:NTF a {
      #1., \@ }%
    {#1., \@}}}
\ExplSyntaxOff

\newcommand{\CASE}[1]{\STATE \textbf{case} #1\textbf{:} \begin{ALC@g}}
\newcommand{\ENDCASE}{\end{ALC@g}}

\newcommand{\DEFAULT}{\STATE \textbf{default:} \begin{ALC@g}}
\newcommand{\ENDDEFAULT}{\end{ALC@g}}
\newcommand{\DEFAULTLINE}[1]{\STATE \textbf{default:} }
\let\footnotesize\scriptsize

\newsavebox{\supbox}
\newcommand{\bsup}{\begin{lrbox}{\supbox}$\tt\scriptstyle}
\newcommand{\esup}{$\end{lrbox}{}^{\usebox{\supbox}}}
\def\eg{\latinabbrev{e.g}}

\algnewcommand{\LineComment}[1]{\State \(\triangleright\) #1}

\renewcommand{\refname}{REFERENCES}

\definecolor{lightpurple}{rgb}{0.8,0.8,1}
\definecolor{codebg}{RGB}{255,255,255}
\definecolor{commentcolor}{RGB}{11,140,11}

\AtBeginEnvironment{quote}{\singlespacing\small\textit}

\colorlet{shadecolor}{gray!10}
\colorlet{framecolor}{black}
\newenvironment{frshaded}{%
	\MakeFramed {\FrameRestore}}%
{\endMakeFramed}

\begin{document}
%


\title{Why are Some Bugs Non-Reproducible?~\\
\Large –An Empirical Investigation using Data Fusion–\vspace{-.3cm}}

\author{\IEEEauthorblockN{Mohammad Masudur Rahman  ~~~ Foutse Khomh  ~~~ Marco Castelluccio$^\dagger$}
	\IEEEauthorblockA{Polytechnique Montreal, Canada, Mozilla Corporation$^\dagger$\\
		\{masud.rahman, foutse.khomh\}@polymtl.ca, mcastelluccio@mozilla.com$^\dagger$}
}
\maketitle

\begin{abstract}
	Software developers attempt to reproduce software bugs 
	to understand their erroneous behaviours and to fix them. 
	Unfortunately, they often fail to reproduce (or fix) them, which leads to 
	faulty, unreliable software systems. However, to date, only a little research has been done to better understand what makes the software bugs non-reproducible. In this paper, we conduct a multimodal study to better understand the non-reproducibility of software bugs. First, we perform an empirical study using 576 non-reproducible bug reports from two popular software systems (Firefox, Eclipse) and identify 11 key factors that might lead a reported bug to non-reproducibility. Second, we conduct a user study involving 13 professional developers where we investigate how the developers cope with non-reproducible bugs. We found that they either close these bugs or solicit for further information, which involves long deliberations and counter-productive manual searches. Third, we offer several actionable insights on how to avoid non-reproducibility (e.g., false-positive bug report detector) and improve reproducibility of the reported bugs (e.g., sandbox for bug reproduction) by combining our analyses from multiple studies (e.g., empirical study, developer study). 
	
\end{abstract}



\IEEEpeerreviewmaketitle

\section{Introduction}\label{sec:introduction}
Software bugs and failures claim trillions of dollars every year. In 2017 alone, 606 software bugs cost the global economy about \$1.7 trillion with 3.7 billion people affected and 314 companies impacted \cite{17T}. Finding such bugs within software code and then fixing them are highly challenging. One of the major challenges is to determine the root cause of a reported bug, which might help better understanding of the bug \cite{parnin}. Developers often attempt to reproduce a bug from its report to determine its root cause and to explore its erroneous behaviours.
Unfortunately, the bug reports often do not contain useful information for reproducing the bugs \cite{missing,oscar-s2r-quality,goodbugreport}.
This leads to (a) an unexpected delay (e.g., three months \cite{worksforme-ubc}) in bug fixing \cite{not-my-bug,predict-reopened} or (b) even worse, the release of a software system without fixing potentially critical bugs, which could be costly in the long run (e.g., Facebook's privacy vulnerability \cite{facebook-vulnerability}). Thus, (a) studying the key factors (a.k.a., characteristics) that could make the reported bugs non-reproducible and
(b) detecting these non-reproducible bugs early in their management cycle -- are open research problems that warrant further investigations. Our work in this paper addresses the first research problem.


\begin{figure*}
	\centering
	\resizebox{6.8in}{!}{%
		\begin{tikzpicture}[scale=1, auto,swap, 	block/.style={
			draw,
			fill=white,
			align=center,
			rectangle, 
			text width={3cm},
			inner sep=3pt,
			fill=lightgray,
			font=\small},
		mylabel/.style={
			font=\small }]
		
		\begin{pgfonlayer}{bg}

		\node[inner sep=0pt] (bug) at (-6.5,2.2)
		{\includegraphics[width=.45in]{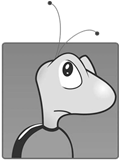}};
		\node[inner sep=0pt] (mozilla) at (-5.2,2)
		{\includegraphics[width=.65in]{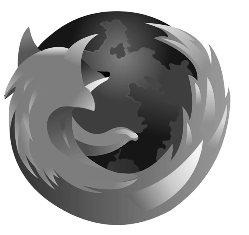}};
		\node at (-5.9,.8) (b) {Firefox};
		\node at (-5.9,.4)(b) {Bug Reports};
		
		\node[inner sep=3pt] at (-4,2) [circle,draw] (1a) {1a};

		\node[inner sep=0pt] (bug) at (-6.5,-1.8)
		{\includegraphics[width=.45in]{img/bug.png}};
		\node[inner sep=0pt] (eclipse) at (-5.2,-2)
		{\includegraphics[width=.65in]{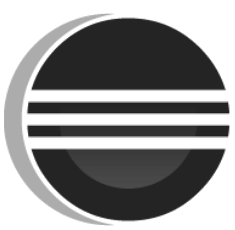}};
		\node at (-5.9,-3.2) (b) {Eclipse};
		\node at (-5.9,-3.6) (b){Bug Reports};
		
		\node[inner sep=3pt] at (-4,-2) [circle,draw] (1b) {1b};
		
		\node[inner sep=0pt] (filter) at (-2,0)
		{\includegraphics[width=.50in]{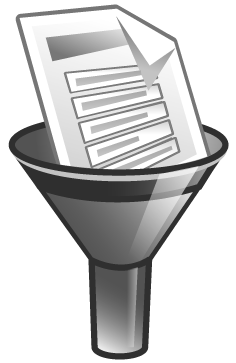}};
		\node[inner sep=0pt] (gt) at (-1.2,0) {\includegraphics[width=.55in]{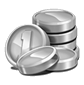}};
		\node at (-1.8,-1.3) {Filtration \& Sampling};
		\node[inner sep=3pt] at (-1.6,1.3) [circle,draw] (2) {2};
		
		\draw[->,thick] (mozilla) -- (filter);
		\draw[->,thick] (eclipse) -- (filter);
		
		\node[inner sep=0pt] (dsbr0) at (1.5,0) {\includegraphics[width=.50in]{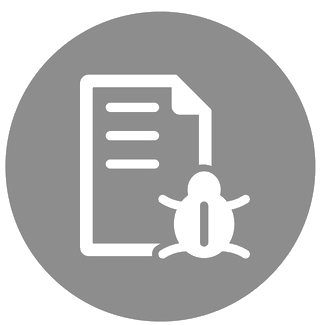}};
		\node[inner sep=0pt] (dsbr) at (2.5,0) {\includegraphics[width=.50in]{img/bf1tr.png}};
		
		\node at (2,1.4) (lab3) {Selected NR};
		\node at (2,1) (lab3) {Bug Reports};
		\node[inner sep=3pt] at (2,-1) [circle,draw] (3) {3};

		\draw[->,thick] (gt) -- (dsbr0);

		
		\node[inner sep=0pt] (setting0) at (5.2,0)
		{\includegraphics[width=.65in]{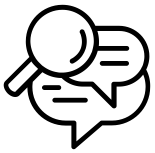}};
		
		\node[inner sep=0pt] (setting) at (6.4,0)
		{\includegraphics[width=.5in]{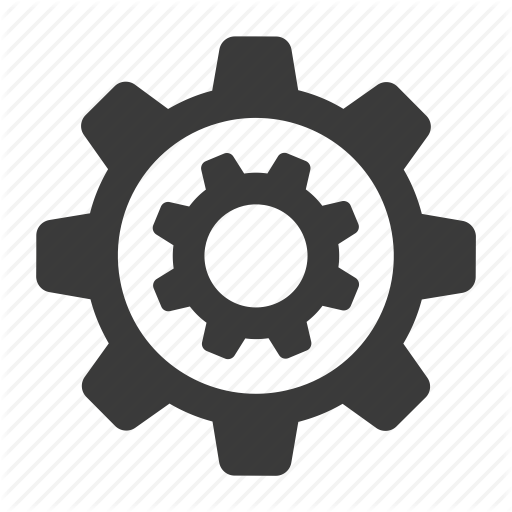}};
		
		\node at (5.6,-1.2) (lab3) {Qualitative Analysis};
		\node at (5.6,-1.6) (lab3) {(Grounded Theory)};
		
		\node[inner sep=3pt] at (5.6,1) [circle,draw] (4) {4};

		\draw[->,thick] (dsbr) -- (setting0);
		
		\node[inner sep=3pt] at (9,3) [circle,draw,fill=lightgray] (f1) {F$_1$};
		\node[inner sep=3pt] at (9,2) [circle,draw,fill=lightgray] (f2) {F$_2$};
		\node[inner sep=3pt] at (9,1) [circle,draw,fill=lightgray] (f3) {F$_3$};
		\node[inner sep=3pt] at (9,0) [circle,draw,fill=lightgray] (f4) {F$_4$};
		\node[inner sep=3pt] at (9,-1) [circle,draw] (b) {.};
		\node[inner sep=3pt] at (9,-2) [circle,draw] (b) {.};
		\node[inner sep=3pt] at (9,-3) [circle,draw,fill=lightgray] (fn) {F$_n$};
		
		\node[inner sep=3pt] at (8,3) [circle,draw] (5) {5};
		\node at (7,1.8)[ellipse, draw=black, fill=gray!10] (rq1) {\normalsize RQ$_1$};

		\draw[->,thick] (setting) -- (f1);
		\draw[->,thick] (setting) -- (f2);
		\draw[->,thick] (setting) -- (f3);
		\draw[->,thick] (setting) -- (f4);
		\draw[->,thick] (setting) -- (fn);
		
		\node at (9,-4) (b) {Key Factors};
		
		\node[inner sep=0pt] (agreement) at (12,0)
		{\includegraphics[width=.60in]{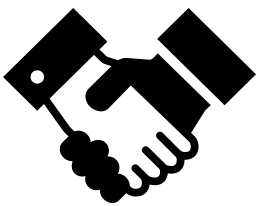}};
		\node at (13.5,.3) (b) {Data};
		\node at (13.5,-.3) (b) {Fusion};
		
		\draw[->,thick] (f1) -- (agreement);
		\draw[->,thick] (f2) -- (agreement);
		\draw[->,thick] (f3) -- (agreement);
		\draw[->,thick] (f4) -- (agreement);
		\draw[->,thick] (fn) -- (agreement);

		\node[block](ubc) at (12,3) {\normalsize \citet{worksforme-ubc}};

		\node[inner sep=0pt] (devstudy) at (12,-3)
		{\includegraphics[width=.65in]{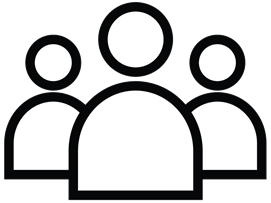}};
		\node at (12.2,-4) (b) {Developer Study};
		\node[inner sep=3pt] at (11.4,-2.2) [circle,draw] (6a) {6a};
		\node at (12.8,-2)[ellipse, draw=black, fill=gray!10] (rq3) {\normalsize RQ$_2$};	
		
		\node at (14.2,-2)[ellipse, draw=black, fill=gray!10] (rq31) {\normalsize RQ$_4$};

		\node at (11.2,1.8)[ellipse, draw=black, fill=gray!10] (rq2) {\normalsize RQ$_3$};
		\node[inner sep=3pt] at (12.6,2) [circle,draw] (6b) {6b};
		
		\node[inner sep=3pt] at (12.6,-.8) [circle,draw] (6c) {6c};
		
		\draw[->,thick] (ubc) -- (agreement);
		\draw[->,thick] (devstudy) -- (agreement);
		
		\node[inner sep=0pt] (summary) at (16,0)
		{\includegraphics[width=.60in]{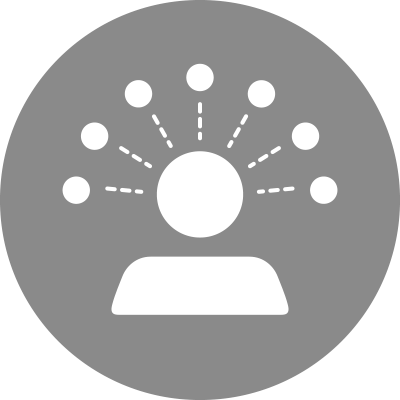}};
		\node at (16.2,-1.4) (b) {Actionable};
		\node at (16.2,-1.8) (b) {Insights};
		\draw[->,thick] (agreement) -- (summary);
		
		\node[inner sep=3pt] at (16,1.1) [circle,draw] (7) {7};
		\node at (15,-.9)[ellipse, draw=black, fill=gray!10] (rq4) {\normalsize RQ$_5$};

		\draw[dashed] (-7.3,-4.3) rectangle (17.2,3.8);

		\end{pgfonlayer}
		\end{tikzpicture}
	}
	\vspace{-.1cm}
	\caption{Schematic diagram of our conducted study}
	\label{fig:sysdiag}
	\vspace{-.3cm}
\end{figure*}   

Existing work from the literature (a) study the characteristics of a good bug report \cite{goodbugreport} and classify the software bugs from open source systems \cite{lintan-emse,bug-class}, (b) predict which bugs get fixed \cite{prediction-fixed,nrfixer}, re-assigned \cite{not-my-bug,bug-field-assignment} or re-opened \cite{predict-reopened} and (c) investigate how bugs are coordinated among various stakeholders (\eg\ software testers, users, developers)  \cite{secret-life-of-bug,bug-repro-collab} and how the misclassification of bugs affects the bug prediction task \cite{its-not-a-bug}. Unfortunately, a little research has been done to better understand what makes the reported bugs non-reproducible. 
\citet{bug-repro-collab} first analyse the social and human aspects of a bug reproduction process with an ethnographic study. Their study explains 
the human collaboration aspect, which might not be enough to explain the complex technical aspect of a bug reproduction process.
\citet{worksforme-ubc} identify several factors (\eg\ Inter-bug dependencies, environmental differences) that might contribute to the non-reproducibility of a reported bug. Although their study sheds some light
on the factors leading to bug non-reproducibility, it does
not go as far as understanding the mitigation strategies currently
implemented in the field. Neither does it examine the mechanisms
to improve the reproducibility of reported bugs. Our work attempts to fill this gap in the literature.


In this paper, we conduct (a) an empirical study to better understand the key factors behind non-reproducibility of software bugs and (b) a developer study to 
understand how the professional developers cope with non-reproducible bugs and how to improve the bug reports.
First, we analyze 576 randomly sampled bug reports marked as \emph{non-reproducible} (250 from Mozilla Firefox + 326 from Eclipse JDT) with a \emph{Grounded Theory} method \cite{grounded-theory} and identify 11 key factors that might lead a reported bug to non-reproducibility.
We also contrast our findings with those from \citet{worksforme-ubc}.
Second, we validate our empirical and analytical findings with a user study involving 13 professional developers (from \texttt{Mozilla} and \texttt{Freelancer}) and gain meaningful insights on how the developers deal with non-reproducible bugs,
which were missing in the earlier work \cite{worksforme-ubc}. Third, by cross-referencing our findings from qualitative analysis and developer study, we report several actionable insights for detecting and then improving the non-reproducible bugs during their submission.
Thus, we answer five important research questions as follows.

\begin{figure}[!t]
	\centering
	\includegraphics[width=3.4in]{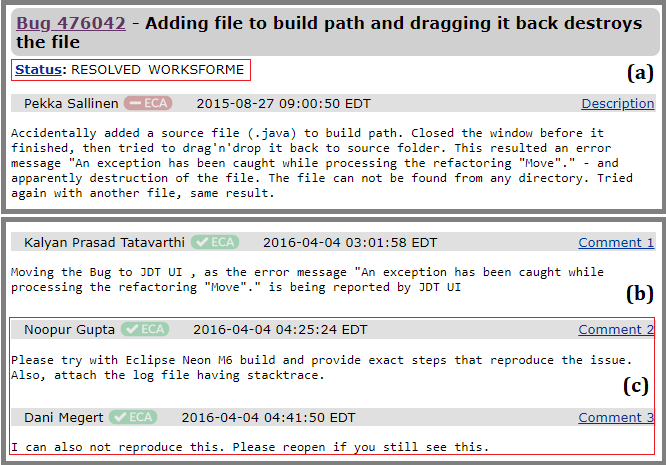}
	\vspace{-.1cm}
	\caption{An example bug reporting thread -- (a) title + description, (b) discussions among stakeholders, and (c) non-reproducibility of the bug}
	\vspace{-.8cm}
	\label{fig:example-bug-report}
\end{figure}

\begin{enumerate}
	\item \textbf{RQ$_1$: What are the key factors that make a reported software bug non-reproducible?}
	 
	\noindent
	To mitigate the issue of bug non-reproducibility, understanding the key causes (or factors) is essential.
	We find 11 key factors (e.g., bug duplication, intermittency, missing information, ambiguous specifications, third-party defects)
	that might lead a reported software bug to non-reproducibility. 
	
	
	\item \textbf{RQ$_2$: What do professional developers consider to be key factors behind the non-reproducibility of bugs?}
	
	\noindent
	Developers' feedback on empirical findings is important to increase confidence in the findings. Our identified key factors were validated by 13 professional developers with an agreement level between 70\% and 90\%.  
	
	
	\item \textbf{RQ$_3$: Do the identified factors match with the relevant earlier findings from the literature}?

	\noindent
	Generalizability of findings across multiple studies and multiple datasets 
	is important to increase confidence in the findings. 
	Five of our identified factors match with the earlier findings derived from a different dataset \cite{worksforme-ubc}. Our study also reveals several novel factors. Thus, our findings not only confirm but also improve the existing understanding of non-reproducible software bugs.
	
	\item \textbf{RQ$_4$: How do the professional developers deal with non-reproducible software bugs?} 
	
	\noindent
	Understanding the current practices for dealing with non-reproducible bugs is important to provide efficient solutions.
	We find that professional developers deliberate over non-reproducible bugs and attempt to collect more information when the bug reports are incomplete or the reported bugs are complex (e.g., intermittent bugs, performance bugs). They also close the duplicate and false-positive bug reports with suitable explanations. 
	Many of these tasks are performed in a counter-productive way due to the lack of appropriate alternatives (e.g., tool supports).


	\item \textbf{RQ$_5$: How to prevent the non-reproducibility and/or improve the reproducibility of reported bugs?}
	
	\noindent 
	Non-reproducibility of software bugs leads to delays in bug-fixing and potentially buggy software releases, which could be mitigated with appropriate tool supports. For example, intelligent tools (1) for detecting duplicate or false-positive bug reports and (2) for improving the software documentations could help avoid non-reproducible bugs. On the other hand, intelligent tools for complementing the incomplete bug reports could improve their reproducibility. Furthermore, sandbox tools where the developers can
	repeat their experiments could be useful for reproducing the complex software bugs (e.g., performance bugs, intermittent bugs).
	
\end{enumerate}



\begin{table*}[!t]
	\centering
	\caption{Study Dataset}\label{table:dataset}
	\vspace{-.2cm}
	\resizebox{6.8in}{!}{%
		\begin{threeparttable}
			\begin{tabular}{l|l|c|c|c|c|c}
				\hline
				\textbf{System} & \textbf{Domain} & \textbf{BTS} & \textbf{Component} & \textbf{Duration} & \textbf{All NR Bugs} & \textbf{Selected NR Bugs}\\
				\hline
				\hline
				Mozilla Firefox & Web browser & Bugzilla & Core & Dec 29, 2017--Dec 29, 2019 & 1,274 & 250\\
				\hline
				Eclipse & Programming IDE & Bugzilla & JDT & Dec 29, 2014--Dec 29 2019 & 326 & 326\\
				\hline
				\textbf{Total} & - & - & - & - & \textbf{1,600} & \textbf{576} \\ 
				\hline
			\end{tabular}
			\centering
			\textbf{NR} = Non-Reproducible, \textbf{BTS} = Bug Tracking System 
		\end{threeparttable}
	}
\vspace{-.4cm}
\end{table*}

\section{Non-Reproducible Bugs}\label{sec:bg}
Software developers often attempt to reproduce their bugs to better understand them. \emph{Non-reproducible} bugs are the ones that cannot be reproduced by using the information found in their corresponding bug reports. They are annotated with several labels such as \emph{``Works on My Machine"}, \emph{``Works For Me"} and \emph{``Cannot Reproduce"} in the bug-tracking systems \cite{worksforme-ubc}. 
About 17\% of the reported bugs could be non-reproducible \cite{worksforme-ubc}, which is a significant amount. Based on a developer study, \citet{goodbugreport} suggest that developers expect at least three components within a bug report -- Observed Behaviour (OB), Expected Behaviour (EB) and Steps to Reproduce (STR). OB describes the erroneous behaviour of a software system whereas EB outlines the correct behaviour of the system. On the other hand, STR provides the steps to reproduce a bug. Bug reports that miss these components could be
difficult to reproduce. According to \citet{missing}, about 65\% of the bug reports miss the correct behaviour (EB) and 49\% of the reports do not contain any steps to reproduce the bugs. Fig. \ref{fig:example-bug-report} shows an example bug that erroneously deletes documents from the build path of an Eclipse project. From the discussions, we also see that the bug report fails to provide any concrete steps or complementary information (e.g., logs, stack traces) for reproducing this bug (Fig. \ref{fig:example-bug-report}-(c)). As a result, the bug was marked as \emph{non-reproducible} (a.k.a., WORKSFORME) by the developers and was abandoned without fixing.


\section{Study Methodology} \label{sec:methodology}
Fig. \ref{fig:sysdiag} shows the schematic diagram of our performed studies in this paper.
We first perform an empirical study on bug non-reproducibility using 576 non-reproducible bugs from two popular software systems. We not only revisit the earlier findings \cite{worksforme-ubc} but also deliver novel insights towards better understanding of non-reproducible bugs. Then we validate our major findings with a developer study and formulate further actionable insights. In this section, we discuss the major steps of our study design as follows.


\subsection{Dataset Construction}\label{sec:dataset}
We use a total of 576 non-reproducible bugs from two popular, mature, open source systems -- Mozilla Firefox and Eclipse -- for our study. 
Table \ref{table:dataset} shows an overview of our study dataset.
Several steps were taken to carefully select these reported bugs. First, we collect all the bug reports from these systems that were marked as WORKSFORME.
Both systems use this tag to mark their \emph{non-reproducible} bugs \cite{wfm-def}. Then we select the ones that were submitted within the last two years from Firefox and within the last five years from Eclipse. We use the recent bugs that the developers could still remember and that can provide timely, relevant, actionable insights. 
We also choose the bug reports concerning two major components -- \emph{Firefox Core} and \emph{Eclipse JDT} -- due to their critical roles. This step provides a total of 1,600 non-reproducible bug reports (1,274 from Firefox + 326 from Eclipse) (Step 1, Fig. \ref{fig:sysdiag}). Second, since manually analysing hundreds of bug reports would be too expensive, we randomly choose 576 bug reports from them. In particular, we identify top 25 critical subcomponents in Firefox Core (e.g., \emph{WebRender}, \emph{Playback}, \emph{JavaScript Engine}) based on their bug report frequency and choose 10 random bug reports from each of these subcomponents, which provides 250 random bug reports. Our sample has a confidence level of 95\% with an error margin of 5.56\%. On the other hand, we choose all 326 bug reports targeting Eclipse from the last five years. Finally, we ended up with 576 (250+326) non-reproducible bugs from two popular software systems (Steps 2-3, Fig. \ref{fig:sysdiag}).

 \subsection{Identifying Key Factors with Qualitative Analysis}\label{sec:gtheory}
 We carefully analyse the information available in the bug reports and attempt to understand the key factors behind the non-reproducibility of their discussed bugs (Steps 4--5, Fig. \ref{fig:sysdiag}). We first establish the scope of our analysis and then employ a widely used qualitative method -- Grounded Theory \cite{grounded-theory} -- for our analysis as follows. 
 
\textbf{Determining the Scope of Manual Analysis:} In modern bug-tracking systems (e.g., Bugzilla), each bug report captures (1) bug description from a reporter and (2)  discussions among various stakeholders (e.g., reporter, developers, testers). Once an encountered bug is reported, the stakeholders engage in a discussion where they attempt to reproduce the bug using the available information at hand. Since we attempt to understand the key issues behind bug non-reproducibility, we analyse both the bug description and the discussion texts from each bug report. While textual contents are mostly prevalent, bug reports might also contain supplementary materials (e.g., stack traces, logs, memory dump, test cases, configuration files) to assist in bug fixing. In this study, we systematically analyse the textual contents and occasionally check the supplementary materials for deriving our insights.
  
\textbf{Grounded Theory:} We analyse bug reports using Grounded Theory method \cite{grounded-theory} to determine the key factors behind bug non-reproducibility. Grounded Theory has been widely adopted in the social science researches to derive theories that are firmly grounded in the data under analysis (\eg\ interview scripts, questionnaires). Recently, this method has also found applications in the Software Engineering researches \cite{bug-fix-thesis-arshad,observed}. We systematically analyse the bug description and discussion texts, and look for potential clues (e.g., missing information, technical difficulty) that might explain the non-reproducibility of a reported bug. Grounded theory method involves three stages of coding as follows.
 
 \textit{(a) Open Coding} consists in breaking the gathered data (e.g., bug reports) into identifiable, interesting chunks. We go through the bug description and discussion texts of each bug report and look for potential clues that might explain the non-reproducibility of a bug. We record our identified clues using a set of key phrases \cite{icsme2020-rep-package}. The core idea was to keep an open mind and to choose as many codes as needed to carefully represent each bug report. In our open coding, 574 unique codes were produced from 576 bug reports. We spent $\approx$100 man-hours in the open coding of 576 bug reports and their discussions.
 
 \textit{(b) Axial Coding} focuses on finding connections among the open codes. In this stage, we place our open codes into a spreadsheet document and annotate the similar or connected codes with the same colours. We consider not only lexical overlap but also semantic relatedness in establishing the connection. Our goal was to divide the open codes into low-level categories. This step provides a set of 33 tentative categories from our 574 open codes above.
 
 \textit{(c) Selective Coding} determines the core variables (or categories) and constructs a theory to explain the phenomenon under study, i.e., the non-reproducibility of the reported bugs. While the axial coding provides low-level categories, we carefully merge them into higher level categories 
 based on their common themes and semantic relatedness.
 This step provides a total of 11 key factors that might explain the non-reproducibility of reported software bugs (Step 5, Fig. \ref{fig:sysdiag}). Each of these key factors is represented using a set of semantically connected key phrases \cite{icsme2020-rep-package}.
 
 \textbf{Determining the Prevalence of Key Factors in Bug Reports:} 
 During open coding, we represent each bug report using a set of suitable key phrases that explain why the bug could not be reproduced. Similarly, each of our identified factors is represented using a set of semantically connected key phrases. To analyse the prevalence of key factors in our dataset, we determine the presence of one or more key factors in each bug report using their overlapping key phrases.
 
 \subsection{Complementing Empirical Study with Developer Study}
 Although our empirical findings are derived from developer discussions, we further validate and complement these findings with a developer study involving 13 professional developers.
 We also investigate how these developers cope with non-reproducible bugs in practice and how the research community could help them.
 We discuss our study setup including questionnaire preparation and participant selection as follows.
  
 \textbf{Questionnaire Preparation:} We first provide a summary of our empirical findings on bug non-reproducibility. Then we ask five different questions on four topics. First, we ask the participants whether they agree or disagree with our identified causes of bug non-reproducibility (Table \ref{table:keyfactors}) using dichotomous questions. Second, we ask how they deal with non-reproducible bugs as a part of their job. Third, we ask how the research community could help the developers in dealing with the non-reproducible bugs. Fourth, we ask the developers about their professional experience level, which was used for the demographic analysis.
 
 
 \textbf{Participant Selection:} We first conduct a pilot study with one professional developer from Mozilla Firefox, which helped us improve our questionnaire. Then we invite the professional developers  from Mozilla, Freelancer and Stack Overflow who have relevant bug-fixing experience
 to answer our questionnaire. We send our invitations to the developers using direct correspondences, organization's mailing lists (e.g., Mozilla Firefox) and public forums (e.g., LinkedIn, Twitter). Thirteen participants responded to our invitation including four developers from Mozilla Firefox. About 23\% of these participants have more than 10 years of professional development experience, 15\% of them have 5 to 10 years of experience and about 39\% have 1 to 5 years of development experience. Our study was non-paid.
 
 \subsection{Analysing Agreement between Independent Findings}\label{sec:agreement}
 Once the key factors behind bug non-reproducibility are identified, we validate them with two independent studies (Steps 6-7, Fig. \ref{fig:sysdiag}). First, we investigate how these factors are assessed by the professional developers.
 We not only determine the severity of each factor but also gain further actionable insights from this developer study. 
 Second, we determine how these factors match with the earlier findings of \citet{worksforme-ubc}. While \citeauthor{worksforme-ubc} adopt an ad hoc method, we employ a systematic method namely \emph{Grounded Theory} \cite{grounded-theory} for the qualitative analysis. We also use a different dataset in this work. Thus, an agreement between our identified factors and the earlier ones would indicate their generalizability 
 and thus also would strengthen the understanding of bug non-reproducibility.

 \section{Study Results}\label{sec:result}
In this section, we present the results of our study by answering the five research questions as follows.

\subsection{RQ$_1$: What are the key factors that make a reported software bug non-reproducible?}
We identify a total of 11 key factors that are likely to explain the non-reproducibility of software bugs. Table \ref{table:keyfactors} shows the identified factors from our qualitative analysis (Section \ref{sec:gtheory}). We explain each of these factors with illustrative examples collected from the bug reports as follows.

\begin{table*}[!t]
	\centering
	\caption{Key Factors behind the Non-Reproducibility of Software Bugs}\label{table:keyfactors}
	\vspace{-.2cm}
	\resizebox{6.8in}{!}{%
		\begin{threeparttable}
			\begin{tabular}{p{1.3in}|p{3.5in} }
				\hline
				\textbf{Key Factor} & \textbf{Overview}   \\  
				\hline
				\hline
				  \emph{F$_1$: Bug Duplication} & The bug might have been already fixed in the recent releases \\
				 \hline
				 \emph{F$_{2}$: False Positive Bug} & The reported issue might not be a bug, but rather indicates a non-existent software feature \\
				 
				 \hline
				  \emph{F$_3$: Bug Intermittency} & The bug does not occur frequently or consistently \\
				\hline
				\emph{F$_4$: Missing Information} & The required information (e.g., steps to reproduce) is missing in the bug report\\
				\hline
				
				  \emph{F$_5$: Ambiguous Specifications} & The expected behaviour of the software application is misunderstood by the reporter 
				   \\
				 \hline	  
				  
				  \emph{F$_6$: Performance Regression} & Performance loss that is encountered as a side effect of recent changes \\
				  \hline
				  
				  \emph{F$_{7}$: Lack of Cooperation} & The reported bug fails to draw the attention of the stakeholders (e.g., developers) \\
				  \hline
				  
				   \emph{F$_8$: Memory Misuse} & The bug is triggered by the mismanagement of memory \\
				  \hline 
				  
				   \emph{F$_9$: Third-Party Defect} & The bug has been triggered by defects in a third-party component\\  
				  \hline
				  
				  \emph{F$_{10}$: Restricted Security Access} & Bugs that warrant specialized authentication or authorization for reproduction \\
				  \hline
				  \emph{F$_{11}$: Touch \& Gestures} & The accessibility bugs that warrant touches, gestures and special interactions \\
				  \hline
				  
			\end{tabular}
			\centering
		\end{threeparttable}
		\vspace{-.4cm}
	}
\vspace{-.5cm}
\end{table*}

\textbf{Bug Duplication (F$_1$)} is one of the key factors behind the non-reproducibility of software bugs. The duplicate bugs are often known to the developers and thus might have been already fixed in recent releases. As a result, they cannot be reproduced with the up-to-date version of the software system. The following discussion comment from Firefox (Bug \#1428773) 
refers to duplicate bugs and explains why it cannot be reproduced.
\begin{quotation}
	\noindent
	\emph{D: ``This looks a lot like the issues in bug 1420748 and related bugs, so it might be fixed by the WR update in bug 1426116."} 
\end{quotation}
Since the duplicate bugs might already be fixed, they are often resolved by closing them as \emph{duplicates} and then pointing their reporters to a recent software version that contains the fixes.

\textbf{False Positive Bugs (F$_{2}$)} (a.k.a., feature requests) might not be reproducible. Software users often discuss about non-existent features in their bug reports; this cannot be reproduced. Sometimes, they also report configuration issues as bugs, which can be resolved with simple tweaking rather than code level changes. For example, the following comment from Firefox (Bug \#1444194) shows how a simple configuration tweak can resolve a bug regarding slow network proxy.  
\begin{quotation}
	\noindent
	\emph{D: ``What happens if you set the about:config pref "security.OCSP.enabled" to 0?"}  ~\\
	\emph{R: ``yep disabling the OCSP check fixed it. Then this works fine .."} 
\end{quotation} 


\textbf{Bug Intermittency (F$_3$)} is another key factor behind bug non-reproducibility. Intermittent bugs have non-deterministic properties and they do not occur frequently or consistently. Thus, they are difficult to reproduce and fix. The following discussion comment from 
Eclipse (Bug \#501488) explains the intermittency of a reported bug. 
\begin{quotation}
	\noindent
	\emph{R: ``I forgot to write that the problem does not always appear, sometimes working as it should. I have inserted a screenshot of the problem."} 
\end{quotation}

\textbf{Missing Information (F$_4$)} is a key problem that makes a reported bug non-reproducible. Developers often look for relevant items in a bug report (e.g., steps to reproduce, stack traces, performance profiles, screenshot, test cases, system configurations) that could help them in reproducing the bug. Unfortunately, in practice, these items often are either missing or not reported carefully. 
The example comments from Eclipse (Bugs \#476042, \#477898) request for missing information in the bug report.
\begin{quotation}
	\noindent
	\emph{D: ``Please try with Eclipse Neon M6 build and provide exact steps that reproduce the issue. Also, attach the log file having stacktrace."} 
\end{quotation}
\begin{quotation}
	\noindent
	\emph{D: ``Please provide the code snippet that reproduces the issue."} 
\end{quotation}
Sometimes, the mere presence of the required items might not be sufficient. The reported bugs could also not be reproduced if the provided information is incomplete or inaccurate. 

\textbf{Ambiguous Specifications (F$_5$)} often lead the reported bugs to non-reproducibility. Bug reporters might misunderstand the expected behaviours of a software system if the specifications are not clearly defined. As a consequence, they might characterize a legitimate functionality as a bug, which could introduce confusion or disagreement during bug reproduction. For example, the following discussion comments from Firefox (Bug \#1477421) indicate a misunderstanding of the specifications regarding autoplay for muted videos.
\begin{quotation}
	\noindent
	\emph{R: ``Actual results: Video autoplays although the sound is muted. Expected results: Video should not autoplay."} 
\end{quotation}
\begin{quotation}
	\noindent
	\emph{D: ``Muted videos are expected to autoplay. This is a design choice."} 
\end{quotation}

\textbf{Performance Regression (F$_6$)} related bugs (a.k.a., performance bugs) are hard to reproduce. They are often subtle and subjective, or dependent on specific characteristics of a machine, which introduces confusion and disagreement among the stakeholders (e.g., reporters, developers) during bug reproduction. For example, the following comment from Firefox (Bug \#1485402) 
indicates the subtle and subjective natures of the performance bug. 
\begin{quotation}
	\noindent
	\emph{D: ``On my 24-core desktop machine, I'm seeing Firefox Nightly 63 being quite a bit *faster* than Chrome DevEdition 70.  I wonder why I'm seeing the opposite of Stephen (reporter)..."} 
\end{quotation}

\textbf{Lack of Cooperation (F$_{7}$)} among the stakeholders is another key factor that could make a reported bug non-reproducible. An earlier study \cite{bug-repro-collab} also suggests that collaboration dynamics could play a major role during bug reproduction. 
Sometimes the reported bugs fail to draw the attention of human developers. They are closed by either the automated bots (e.g., Eclipse Genie) or the reporters
and are marked as WORKSFORME.
Bug reproduction might also fail due to the lack of response from the reporters. For example, the following comment from Eclipse (Bug \#495568) indicates that the bug cannot be reproduced due to the lack of cooperation (and required information) from the bug reporter.
\begin{quotation}
	\noindent
	\emph{D: ``No further feedback, closing. Please reopen if you can confirm the problem and provide reproducible examples."} 
\end{quotation}


\textbf{Memory Misuse (F$_8$)} such as \emph{memory leaks}, \emph{memory overflows}, and \emph{concurrent modifications} might trigger the complex bugs that are difficult to reproduce.
A leak of a small object that is hardly noticeable during the execution of the program might cause the memory usage to grow unbounded. Such issues could also be compounded by legacy hardware.
For example, the following comment from Firefox (Bug \#1547586) indicates the complex nature of a memory related bug regarding excessive RAM usage.

\begin{quotation}
	\noindent
	\emph{D: ``I am unable to reproduce. I created a new profile, opened ... until the page was finished loading, measured RAM ... and a Firefox about:memory report. Then I disabled accessibility services, and restarted with ..., and re-measured RAM. I did not see any significant change. I tried this on a 9 year old laptop and a 2 year old laptop and saw no significant differences."} 
\end{quotation}

\textbf{Third-Party Defects (F$_9$)} are often responsible for non-reproducible bugs. Modern software systems are routinely developed with third-party dependencies (e.g., libraries, resources) and environmental specifications (e.g., OS, memory, hardware, plug-ins, anti-viruses \cite{dll-injection}) that might trigger bugs and failures. However, these bugs might not be reproducible since the developers often do not have enough control over them, or do not have a way to install the same third-party software which is the root cause of the bug. For example, the following comment from Firefox (Bug \#1427890) suggests that the bug could be specific to an operating system.
\begin{quotation}
	\noindent
	\emph{D: ``It works for me on Firefox 57 with windows 10. Since reporters use windows 7, maybe it is related to the windows version."} 
\end{quotation}

\textbf{Restricted Security Access (F$_{10}$)} is another important factor behind the non-reproducibility of software bugs.
Although bug reports are supposed to provide the required information for reproducing the bugs, many confidential items (e.g., user credentials, security certificates) cannot be shared publicly. Thus, reproducing the end-user's experience accurately could be challenging. For example, 
if there is a bug with Firefox on Netflix and the developer does not have a NetFlix account, then she might not be able to reproduce it.
The following comment from Firefox (Bug \#1594272) indicates the non-reproducibility of a bug due to possibly restricted security access.
\begin{quotation}
	\noindent
	\emph{D: ``Hi Mark, I wasn't able to reproduce the bug since I don't have an account but I've chosen a component for this bug ..."} ~\\
\end{quotation}

\textbf{Touch \& Gestures (F$_{11}$)} are often hard to imitate precisely, which could make accessibility-related bugs non-reproducible. For example, the following comment from Firefox (Bug \#1457726) discusses the challenges in reproducing a touch/gesture related bug.
\begin{quotation}
	\noindent
	\emph{D: ``I have tested this issue on a Surface machine with Windows 10 x64 ... and haven't managed to reproduce the issue. After opening multiple tabs and tapping on the "x" close button, the tab is automatically closed"} 
\end{quotation}
\FrameSep3pt
\begin{frshaded}
	\noindent
	\textbf{Summary of RQ$_1$-(a):} There are at least \textbf{11 key factors} (e.g., \emph{bug duplication, intermittency, missing information, ambiguous specifications, third-party defects}) that could lead a reported software bug to non-reproducibility. 
\end{frshaded}


To better understand the importance of each of the identified factors, we analyse their prevalence in our dataset of non-reproducible bugs (Section \ref{sec:gtheory}).
We determine the presence of one or more key factors in each bug report and then summarize our findings (Table \ref{table:factor-stat}) as follows.

\begin{table*}[!t]
	\centering
	\caption{Prevalence of the Key Factors in Non-Reproducible Bugs}\label{table:factor-stat}
	\vspace{-.1cm}
	\resizebox{5.5in}{!}{%
		\begin{threeparttable}
			\begin{tabular}{l|c|c||l|c}
				\hline
				 \multicolumn{3}{c||}{\textbf{Proposed Study}} & \multicolumn{2}{c}{\textbf{\citeauthor{worksforme-ubc}}}\\  
				\hline
				\textbf{Key Factor} & \textbf{Firefox} & \textbf{Eclipse} & \textbf{Key Category} & \textbf{All} \\
				\hline
				\hline
				F$_1$: Bug Duplication & 	\textbf{26.83}\% & \textbf{31.33}\% & C$_1$: Inter-bug Dependencies & \textbf{45.00}\% \\
				\hline
				F$_{2}$: False Positive Bug &	\textbf{4.57}\% & \textbf{21.67}\% & - & - \\
				\hline
				F$_3$: Bug Intermittency &	\textbf{26.22}\% & 2.61\% & C$_5$: Non-deterministic Behaviour & 3.00\% \\
				\hline
				F$_4$: Missing Information &	1.52\% & \textbf{13.84}\% & C$_3$: Insufficient Information & \textbf{14.00}\% \\
				\hline
				F$_5$: Ambiguous Specifications &	\textbf{5.18}\% & \textbf{9.40}\% & C$_4$: Conflicting Expectations & \textbf{12.00}\% \\
				\hline
				F$_6$: Performance Regression &	\textbf{8.54}\% & 1.83\%  & - &- \\
				\hline
				F$_{7}$: Lack of Cooperation &   3.96\% & 3.66\% & - & - \\
				\hline
				F$_8$: Memory Misuse &	4.88\% & 1.00\% & - & - \\
				\hline
				F$_9$: Third-Party Defects &	1.83\% & 2.35\% & C$_2$: Environmental Differences & \textbf{24.00}\% \\ 
				\hline
				
				F$_{10}$: Restricted Security Access & 4.27\% & 0.00\% & - & - \\
				\hline
				F$_{11}$: Touch \& Gestures &	2.44\% & 0.00\% & - & - \\
				\hline
				\textbf{Miscellaneous} & 9.76\% & 12.53\% & C$_6$: Others & 2.00\% \\
				\hline 
			\end{tabular}
		\end{threeparttable}

	}
\vspace{-.3cm}
\end{table*}

Table \ref{table:factor-stat} shows the prevalence of 11 key factors in our dataset.
We see that bug duplication is a major factor behind bug non-reproducibility. About 29\% of bugs from the dataset cannot be reproduced since they are duplicates and were possibly fixed earlier. Both Firefox and Eclipse systems encounter a significant number of duplicate, non-reproducible bugs (e.g., 27\%--31\%). Bug intermittency is another prevailing factor behind the bug non-reproducibility. On average, 14\% of the bugs do not occur frequently and consistently, which makes it hard to reproduce them. Up to 26\% of the Firefox bugs are intermittent in nature. 
Developers also fail to reproduce at least 8\% of the bugs due to missing information (e.g., steps to reproduce, stack traces, test cases). This problem is especially severe for Eclipse where $\approx$14\% of the bug reports lack the required information for reproduction. During our analysis, we also note that the mere presence of items might not be sufficient rather they should be complete and accurate.
Ambiguous specification is another important factor that could lead $\approx$8\% of bugs to non-reproducibility. That is, the expected behaviours were either ill-defined or outdated, and the users considered the legitimate software behaviours as bugs. 
From Table \ref{table:factor-stat}, we also see that performance regression and false-positive bugs are also two important factors behind the non-reproducibility of bugs. Minor performance losses as a side effect of recovery from the critical bugs are often acceptable to the developers. 
Hence, they might be reluctant to reproduce these performance bugs. On the other hand, in false-positive bug reports, the reporters complain about non-existent software features, which are impossible to reproduce. The remaining key factors (e.g., third-party defects, memory misuse, restricted security access) lead $\approx$12\% of the bugs to non-reproducibility, which is also a significant amount. Finally, $\approx$11\% of the bugs from our dataset 
are application-specific (e.g., video player autoplay problem, refactoring failure) that  
cannot be reproduced due to miscellaneous reasons (e.g., novice mistakes). 


We also investigate how one or more key factors might lead the software bugs to non-reproducibility. We found that 75\% of the non-reproducible bugs from our dataset cannot be reproduced due to one of the key reasons (e.g., bug duplication, intermittency, missing information). On the other hand, about 22\% of the bugs have two key factors and 2\% of the bugs have three factors behind their non-reproducibility. 
Bug non-reproducibility due to multiple factors might be more difficult to resolve than that due to single factor.
\FrameSep3pt
\begin{frshaded}
\noindent
\textbf{Summary of RQ$_1$-(b):} About \textbf{75\%} of the selected bugs are non-reproducible because of single key factors whereas the remaining ones are made non-reproducible by a combination of two or more key factors (e.g., \emph{intermittency + regression}).
\end{frshaded}
\vspace{-.2cm}

\begin{figure}[!t]
	\centering
	\includegraphics[width=3.15in]{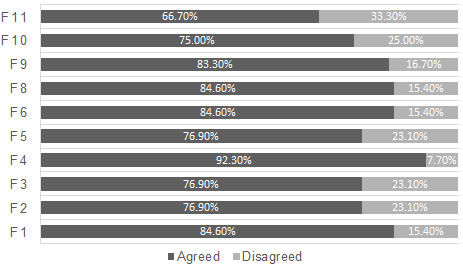}
	\vspace{-.2cm}
	\caption{Developers' responses on the key factors behind bug non-reproducibility}
	\vspace{-.6cm}
	\label{fig:agreed-disagreed}
\end{figure}

\subsection{RQ$_2$: What do the professional developers consider to be key factors behind the non-reproducibility of bugs?}\label{sec:rq1c}
In the developer study, we present our identified factors behind bug non-reproducibility (Table \ref{table:factor-stat}) to the professional developers.
We collect their responses on whether they agree or disagree with these factors. Fig. \ref{fig:agreed-disagreed} shows that about 92\% of the participants consider missing information (F$_4$) to be the major cause of bug non-reproducibility. About 85\% of the participants (i.e., 11 of 13) agree that duplicate bugs (F$_1$), performance bugs (F$_6$), memory misuse related bugs (F$_8$), and third-party defects (F$_9$) are often hard to reproduce. According to $\approx$75\% of the professional developers (i.e., 10 of 13), false-positive bug reports (F$_2$), bug intermittency (F$_3$), ambiguous software specifications (F$_5$), and restricted security access (F$_{10}$) could lead the reported bugs to non-reproducibility. Finally, 67\% of the developers (i.e., 9 of 13) agree that touch and gesture related bugs (F$_{11}$) could also be difficult to reproduce due to their subtle, subjective nature.    
 All these findings above suggest that the professional developers agree to a large extent with our identified factors behind the non-reproducibility of bugs.

We also provide free-form text boxes in our questionnaire to allow the developers to mention any factor that was not included in our list.
Two more important causes were identified from their responses. First, hardware faults are very hard to reproduce. These bugs might need a specific combination of hardware and software (e.g., device drivers) and a long set of steps to reproduce. They also align with one of our factors-- \emph{third-party defects} (F$_9$).  
Second, bugs connected to a random function that is initialized with an unknown seed could also be hard to reproduce due to their non-deterministic nature.
\FrameSep3pt
\begin{frshaded}
	\noindent
	\textbf{Summary of RQ$_2$:} About \textbf{70}\%--\textbf{90}\% of the professional developers agree with the factors behind bug non-reproducibility derived from our empirical study. They also point out two additional types of bugs (e.g., hardware faults, random function bugs) that are difficult to reproduce.
\end{frshaded}


\subsection{RQ$_3$: Do the identified factors match with the earlier findings from the literature?} \label{sec:rq2}
Our first research question (RQ$_1$) identifies a list of key factors behind bug non-reproducibility (Table \ref{table:keyfactors}) using qualitative analysis, which were cross-examined by a group of professional developers (RQ$_2$). 
However, the generalizability of bug non-reproducibility could be further strengthened by validating these factors against the previously reported causes \cite{worksforme-ubc}. \citet{worksforme-ubc} report six major causes of bug non-reproducibility. Table \ref{table:factor-stat} shows their reported causes. We analyse each of our key factors, identify the semantically equivalent causes from their list by consulting their corresponding explanations and examples, and then determine the agreement between these two lists as follows.  


From Table \ref{table:factor-stat}, we see 
that five of our key factors can be comfortably mapped to their Top-5 causes as follows. First, both their study and ours suggest that bug duplication (i.e., F$_1$$\leftrightarrow$C$_1$) is the most dominant factor behind bug non-reproducibility. That is, a significant number of non-reproducible bugs are already fixed. 
Second, while bug intermittency (i.e., F$_3$$\leftrightarrow$C$_5$) is another important factor according to our analysis, \citeauthor{worksforme-ubc} found it less important. 
Third, both studies agree that missing information (i.e., F$_4$$\leftrightarrow$C$_3$) is a major factor that could lead $\approx$14\% of the reported bugs to non-reproducibility. 
Fourth, ambiguous specifications semantically match with conflicting expectations (i.e., F$_5$$\leftrightarrow$C$_4$) due to misunderstanding of the software's correct behaviours. Both their study and ours report this as an important cause
behind bug non-reproducibility. 
Fifth, third-party defects and environmental differences could also be considered as equivalent causes of bug non-reproducibility (i.e., F$_9$$\leftrightarrow$C$_2$). The environmental differences are mostly created by the third-party items (e.g., operating system, network configurations) and the software bugs triggered by them could be hard to reproduce for the developers. Thus, in essence, our study reproduces all the major causes reported by \citeauthor{worksforme-ubc}, which strengthens the generalizability of our findings. 

Besides the existing equivalent causes, we also identify several novel causes of bug non-reproducibility that were not previously known. For example, we found that false-positive bug reports could be a major source of non-reproducibility (e.g., F$_2$, 14\% bugs). 
We also found that performance bugs could be difficult to reproduce since they are often subjective in nature. Minor performance loss as a side effect of critical changes is often overlooked by the developers. According to our analysis, memory misuse related bugs (e.g., memory leak, memory overflow) are also hard to reproduce. 
We also found three other factors -- lack of cooperation, restricted security access, touch \& gestures -- that could be responsible for 7\% of non-reproducible bugs. 
Furthermore, our developer study reveals two more bugs (e.g., hardware faults, bugs connected to random function) that could be very hard to reproduce.

We also compare our findings with the earlier ones \cite{worksforme-ubc} in terms of assigned cause categories. According to \citeauthor{worksforme-ubc}, each bug report could be non-reproducible due to only one major cause. However, we found that at least 25\% of the reported bugs could be non-reproducible because of a combination of two or more factors. 
\begin{frshaded}
	\noindent
	\textbf{Summary of RQ$_3$:} \textbf{Five} of our key factors match with the previously reported causes of bug non-reproducibility \cite{worksforme-ubc}. Our study also reports \textbf{seven novel factors} (false-positive bugs, performance regression, memory misuse, restricted security access, touch \& gestures, hardware faults, bugs from random functions) that could lead the reported bugs to non-reproducibility. Thus, our study strengthens the understanding of bug non-reproducibility both by confirming the earlier findings and by uncovering new factors. 
\end{frshaded}




\subsection{RQ$_4$: How do the professional developers deal with non-reproducible bugs in practice?}
In our developer survey, we ask the developers about how they handle the non-reproducible bugs in practice. We wanted to know what actions they take when the reported bugs cannot be reproduced due to various causes (e.g., Table \ref{table:keyfactors}).
We carefully analyze their qualitative responses against our questions, detect the general themes, and then summarize their actions in respect to the non-reproducible bugs as follows.

\textbf{(a) Duplicate, non-reproducible software bugs are generally closed by the developers}. 
That is, if the developers find a duplicate bug to be working in the latest release, they might close it as WORKSFORME. 
They could also try to find whether the patch of the original bug solves the bug at hand, and
then mark this bug as a DUPLICATE of the original bug.


\textbf{(b) Developers generally close the false-positive bugs as INVALID}. 
They also consult with official functional specifications and occasionally send an explanation to the bug reporters. For example, according to one participant, \emph{``I write a comment explaining why it is a false positive and then close the bug."}

\textbf{(c) Developers attempt to collect useful information from various sources when they encounter intermittent bugs}. For example, if the bug leads to a system crash, they ask for crash dump from the reporter. They also look for debugging logs or system logs associated with the bug, which can provide them rich contexts or insights.
They also search for fellow developers and testers who might have experience with similar bugs, and then delegate the reproduction task to them. The intermittent bugs are also marked as low priority bugs by the developers. That is, if they are not encountered again for a certain period (e.g., 12 weeks), they are eventually closed.

\textbf{(d) Developers request for more information (e.g., steps-to-reproduce, screencast) from the reporters when the bugs cannot be reproduced due to missing information}. If the information is not provided in a timely fashion (e.g., two weeks), then the bug is closed as WONTFIX.

\textbf{(e) Developers ask for further clarifications from the reporters when the bugs cannot be reproduced due to conflicting expectations}. In particular, they explain the expected outcome of a software based on official functional requirements, request for the clarifications and then close the bug if no feedback is received within a certain period (e.g., two weeks).

\textbf{(f) Developers request for performance tracing information (e.g., performance profiles) when they deal with performance bugs}. Such information might help them identify the source of performance bottleneck. They also look for the colleagues who might have relevant expertise.

\textbf{(g) Developers carefully examine the third-party dependencies, their versions and compatibility} when they encounter non-reproducible software bugs triggered by third-party defects. They also check the logs for potential clues and use docker containers for more in-depth investigation. While they emphasize on using only authentic, well-tested plug-ins, many of them are in favour of banning such third-party components that have a strong negative impact upon the main applications (e.g., Firefox browser).

\textbf{(h) Non-reproducibility of software bugs due to restricted security access is a major concern for the developers}. They often work with the reporters closely to help her debug and potentially fix the bug. They also request for regression range from the reporter that might have induced a bug. They might also collect appropriate permission and dummy accounts from the testers to reproduce the bugs.

\textbf{(i) Developers attempt to optimize their code when they deal with memory misuse related bugs.} They increase the memory size for their application and perform extensive debugging to avoid any potential memory leaks. Although the developers claim that they hardly encounter touch/gesture related bugs, they want to take help from the experts in dealing with these bugs.
\begin{frshaded}
	\noindent
	\textbf{Summary of RQ$_4$:} Developers manually identify and close the duplicate and false-positive bug reports. They often look for useful, complementary information from multiple sources when they deal with complex bugs such as intermittent bugs, performance bugs or third-party bugs. They also work closely with various stakeholders (e.g., fellow developers, testers, reporters) and often delegate bug reproduction task to them.
\end{frshaded}


\subsection{RQ$_5$: How to prevent the non-reproducibility and/or improve the reproducibility of reported bugs?}
In our developer study, we ask the developers about how the research community might be able to assist them in dealing with non-reproducible bugs. Given their responses and our empirical analysis,
we provide a list of actionable insights both 
for \emph{preventing} the non-reproducibility and
for \emph{improving} the reproducibility of software bugs as follows.

\textbf{(a) Develop intelligent tools for detecting the duplicate bugs.} About 29\% of the non-reproducible bugs are duplicate bugs, which are already fixed (e.g., Table \ref{table:factor-stat}). Most of these bugs are marked as duplicates by the developers during their failed attempts for reproduction. These reproduction efforts could be saved by carefully detecting the duplicate bugs before their submission. One of our study participants responds, \emph{``Help finding duplicate bugs automatically."}
Unfortunately, many existing tools for detecting the duplicate bugs might not be mature enough for practical use. In particular, they simply rely on textual features \cite{oscar-duplicate,lo-duplicate}, meta data from bug reports (e.g., products, components) \cite{duplicate-search-based} or execution traces \cite{duplicate-bug-nl-exec} for detecting the duplicate bugs,  
 and as a consequence, might fail to detect the complex duplicate bugs that have different symptoms but share the same root causes. Therefore, intelligent tools or techniques are warranted that can accurately detect the duplicate bugs during their report submission and thus can save the wasted efforts in failed reproduction. Furthermore, by putting together multiple similar bugs that share the same root cause, such a tool might equip the developers with enough information for a single bug. One of our developer participants also confirms --\emph{``The more information, the better."}  
 

\textbf{(b) Develop intelligent tools for detecting the false-positive bug reports.} About 5\%--22\% of the non-reproducible bugs are false-positive bugs. They often discuss the non-existent features of a software system that can be neither exercised nor reproduced by the developers. However, this non-reproducibility is 
detected by the developers during their failed reproduction attempts and deliberations, which could be costly. Thus, intelligent tools that can detect the false-positive bug reports during the submission could save valuable development time and efforts. A few existing technique \cite{bug-feat-cls,bug-feat-cls-maalej} attempt to separate the bug reports from the feature requests by analysing their textual features, which might always not be enough. In particular, the underlying semantics could be crucial to separate the software bugs from the features. Thus, further research is warranted to prevent the submission of feature requests as bug reports in the bug-tracking system.   


\textbf{(c) Complement the incomplete bug reports.} Bug reports often lack the elements that are crucial to bug reproduction (e.g., steps to reproduce, expected behaviour, stack traces) \cite{missing}. 
 A few studies \cite{recdroid,oscar-s2r-quality} attempt to reproduce the reported bugs by constructing appropriate test cases from the available information in the bug reports (e.g., steps to reproduce). Unfortunately, they are not sufficient since they are likely to fail when the bug reports lack the required information.
Thus, more intelligent tools and techniques are warranted that (1) can help the 
reporters improve their bug reports during submission or (2) can automatically complement the incomplete bug reports by leveraging the historical information. 
For example, incomplete bug reports could be complemented with partial but valuable information collected from their duplicate or similar bug reports (e.g., stack traces, screen shots).


\textbf{(d) Improve software specifications and documentations.} A significant fraction of the reported bugs (e.g., 8\%) cannot be reproduced due to conflicting expectations between the reporters and the developers. Such a conflict is often triggered by an incorrect understanding of the expected behaviours of a software system, which underscores the need for up-to-date, readable software specifications.
One of our study participants responds --\emph{``I do see that some companies' documentations are vague or ambiguous, so developers, QA, managers, or users may not have a clear understanding on the requirement."} There have been a few tools for creating software documentations from the code (e.g., Doxygen \cite{doxygen}, srcML \cite{srcml}). Since they provide API-level documentations, they might be useful to the developers but not to the users of a software who need more high-level documentations.
Thus, further research is warranted on (1) how to validate the correctness of existing software documentations, (2) how to improve the poor-quality software documentations, and (3) developing tools and techniques that can offer suitable, high-level documentations for software users. Tools that can point the users to the right location within the software documentations could also be very useful.

\textbf{(e) Develop appropriate sandbox to assist in the bug reproduction.} To investigate several complex bugs (e.g., intermittent bugs, concurrency bugs, performance bugs), software systems need to be executed repeatedly. For example, intermittent bugs are non-deterministic and their true characteristics could only be understood from multiple executions. Developers might need to contrast between a normal execution and a crash using their memory dumps when they deal with memory/concurrency bugs. They might also need to compare among the performance profiles from multiple executions to identify the performance bugs. According to the developers, there is a marked lack of such tools and technologies that could help them execute their software applications repeatedly and reproduce these complex bugs. For example, one of our developer participants responds, \emph{``It would be interesting to have a tool that allows the run of a task multiple times and reporting some relevant information as memory usage, dependencies errors, etc."} There have been a few relevant tools (e.g., Firefox Profiler \cite{ff-profiler}, rr \cite{rr}, Pernosco \cite{pernosco}). \emph{Firefox Profiler} can analyse the performance profiles of Firefox and the Gecko browser engines. On the other hand, \emph{rr} and \emph{Pernosco} can record program executions during testing, which could be useful for debugging. However, many of these tools might be limited in scope (i.e., Firefox-specific) or not well-adapted for reproducing bugs. Thus, further research is warranted to 
come up with an appropriate sandbox for reproducing the bugs.

%
%


\textbf{(f) Find the people with right expertise automatically.} Software developers often look for fellow developers and testers with relevant expertise during reproducing complex bugs (e.g., intermittent bugs, concurrency bugs, performance bugs). Although the search might be trivial for a small development group, it could be a major challenge for a geographically distributed, large group. Besides, the relevant expertise might not be obvious and could be hidden as low-level code changes within the version control history. Thus, an intelligent tool for finding the right people might greatly help the developers. There have been a rich literature on finding experts during bug triaging \cite{triage-survey}. Many of these studies simply rely on the texts of a bug report rather than its semantics (e.g., bug types, root causes) to find similar past bug reports and then suggest their assigned developers as experts, which might not be effective enough for practical, widespread adoption. Many existing techniques also rely on naive heuristics (e.g., commit history, code churn) as a proxy of developer's expertise \cite{correct,pick}, which might not be enough. Thus, developing more effective tools for finding the experts during bug reproduction could be a scope for future research.
\begin{frshaded}
	\noindent
	\textbf{Summary of RQ$_5$:} Software developers need intelligent, effective tools for (1) detecting the duplicate or false-positive bug reports, (2) complementing the incomplete bug reports, (3) improving the software documentations, and (4) finding the people with right expertise. They also need a sandbox tool where they can repeat their experiments as a part of reproducing the complex bugs (e.g., performance bugs) and exploring the execution space of their software systems.
\end{frshaded}

\section{Threats to Validity} \label{sec:threat}
Threats to \emph{internal validity} relate to experimental errors and biases \cite{wordsim}. Our key factors behind bug non-reproducibility were derived from a qualitative study (Section \ref{sec:gtheory}), which could be a source of subjective bias. However, our identified factors were validated by a group of 13 professional developers with an agreement level of 70\%--90\% (RQ$_2$). Hence, such a threat might be mitigated. 
Developers might sometimes use WORKSFORME tag loosely or inconsistently, which might introduce some noise in our dataset \cite{worksforme-ubc}. However, since we carefully analyse each of the 576 bug reports and finally summarize our findings, the impacts of such noise might not be significant.

Threats to \emph{external validity} relate to generalizability of our findings \cite{wordsim}. We analyse 576 bug reports from two open source systems (Firefox and Eclipse), which might not be representatives for the proprietary software systems. However, our findings align with that of an earlier study \cite{worksforme-ubc} performed using a different dataset (open source + proprietary) (RQ$_3$), which possibly indicates the generalizability of our study.

The observations from our developer study and our conclusions drawn from them could be a source of threat to \emph{conclusion validity} \cite{emse2018masud}. In particular, there could be a few unseen variables behind the non-reproducibility of bugs (e.g., developer's inexperience, technical infeasibility), which might have been overlooked accidentally. However, we share our dataset \cite{icsme2020-rep-package} publicly for third-party replication and reuse.

\section{Related Work} \label{sec:related}
There have been several studies that analyse the characteristics of a good bug report \cite{goodbugreport} or classify the software bugs from open source systems \cite{lintan-emse,bug-class}. Many studies attempt to predict which bugs get fixed \cite{prediction-fixed,nrfixer}, re-assigned \cite{not-my-bug,bug-field-assignment} or re-opened \cite{predict-reopened}. A few studies also investigate how bugs are coordinated among various stakeholders (\eg\ software testers, users, developers) \cite{secret-life-of-bug,bug-repro-collab} and how the misclassification of bugs affects the bug prediction task \cite{its-not-a-bug}. 
Unfortunately, to date, only a little research has been done to better understand what makes the reported bugs non-reproducible or how to improve their reproducibility during the report submission. 

\citet{worksforme-ubc} first identify six major causes that might explain the non-reproducibility of software bugs (\eg\ Inter-bug dependencies, environmental differences). While their work is a source of inspiration, it does not provide actionable insights on how to detect or improve the non-reproducible bugs during their submission. Their findings were also not validated by the developers.   

\citet{valid-invalid} analyse five different dimensions related to software bugs (e.g., bug report texts, reporter's experience, developer-reporter collaborations) and classify \emph{valid} and \emph{invalid} bug reports using machine learning. Although their work is related to ours, their adopted features might not be appropriate to characterize the non-reproducible bugs. Furthermore, non-reproducibility of the bugs might always not mean that they are invalid bugs.

\citet{bug-repro-collab} analyse social and human aspects of a bug reproduction process with an ethnographic study. Since their findings focus on human collaboration dynamics, they might also not be enough to properly explain the complex technical aspect of a bug reproduction process. 


Unlike many earlier studies above, we conduct an extensive qualitative study with Grounded Theory method \cite{grounded-theory} using 576 bugs reports from Firefox and Eclipse systems, identify 11 key factors behind bug non-reproducibility, and then validate our major findings with 13 professional developers from the industry (e.g., Mozilla). We not only (1) capture 
how the professional developers cope with non-reproducible bugs 
but also (2) offer a list of actionable insights by combining information from multiple analyses (empirical study, developer study), which makes our work novel.
Our findings are generalizable (RQ$_3$) and datasets are publicly available \cite{icsme2020-rep-package}.

\section{Conclusion and Future Work} \label{sec:conclusion} Non-reproducibility of software bugs is a major challenge for the developers since it prevents/delays the bug-fixing. Unfortunately, to date, only a little research has been done to understand the non-reproducibility of bugs. In this paper, we conduct an empirical study using 576 non-reproducible bug reports, and identify 11 key factors behind bug non-reproducibility (e.g., bug duplication, bug intermittency, missing information, false-positive bugs). We not only validate our findings using the feedback from 13 professional developers but also investigate how they cope with non-reproducible bugs. 
Finally, we provide several actionable insights on how to avoid non-reproducibility and/or improve reproducibility of the reported bugs.
By leveraging these insights, future work could focus on developing effective tools and technologies to assist in the bug reproduction (e.g., sandbox for bug reproduction).

\section*{Acknowledgment}
This work was supported by Fonds de Recherche du Quebec (FRQ) and
the Natural Sciences and Engineering Research Council of Canada
(NSERC). We would also like to thank all the anonymous respondents to the survey.

\balance


\bibliographystyle{plainnat}
\setlength{\bibsep}{0pt plus 0.3ex}
\bibliography{sigproc}

\end{document}